\documentclass[a4paper]{scrartcl} %% [twocolumn]
\usepackage{amsmath,amsfonts,amssymb}
\usepackage{graphicx}
\usepackage[left=2cm,right=2cm,top=2cm,bottom=3cm]{geometry} %% ,includeheadfoot
\usepackage{bookman}
\newcommand{\beginsupplement}{%
        \setcounter{table}{0}
        \setcounter{section}{0}
        \renewcommand{\thesection}{S\arabic{section}}%
        \renewcommand{\thetable}{S\arabic{table}}%
        \setcounter{figure}{0}
        \renewcommand{\thefigure}{S\arabic{figure}}%
     }
\usepackage{authblk}
\author[1]{Holger Sennhenn-Reulen}
\affil[1]{\small{Department of Growth and Yield, Northwest German Forest Research Institute, G\"ottingen, Germany}}
\title{Bayesian Regression for a Dirichlet Distributed Response using Stan}
\date{\today\vskip 8mm\small{\textbf{Note: This manuscript is not peer reviewed, but contains many program code sections making the presented results fully reproducible using the statistical software \textsf{R}.}}}
\begin{document}
\maketitle
\paragraph{Abstract} 
For an observed response that is composed by a set -- or vector -- of positive values that sum up to 1, the Dirichlet distribution~\cite{bolshev2018} is a helpful mathematical construction for the quantification of the data-generating mechanics underlying this process. 
In applications, these response-sets are usually denoted as \textsl{proportions}, or \textsl{compositions of proportions}, and by means of covariates, one wishes to manifest the underlying signal -- by changes in the value of these covariates -- leading to differently distributed response compositions. 

This article gives a brief introduction into this class of regression models, and based on a recently developed formulation \cite{maier2014}, illustrates the implementation in the Bayesian inference framework \textsf{Stan}. 

\section{Introduction}
Let us denote by $\mathbf{y}$ a vector of proportion values that is an observation of a random variable~$\mathbf{Y}$. 
As an application example, we can access blood sample composition data via R add-on package \textsf{DirichletReg} \cite{maier2015}.
The data set \texttt{BloodSamples} gives 36 compositions of Alb., Pre-Alb., Glob. A, and Glob. B in relation to two types of diseases (14 patients suffer from disease A, 16 from disease B and 6 are unclassified). 
The first 6 observations show the following values:
\begin{verbatim}
library("DirichletReg")
Bld <- BloodSamples
head(Bld)
   Albumin Pre.Albumin Globulin.A Globulin.B Disease New
A1   0.348       0.197      0.201      0.254       A  No
A2   0.386       0.239      0.141      0.242       A  No
A3   0.471       0.240      0.089      0.200       A  No
A4   0.427       0.245      0.111      0.217       A  No
A5   0.346       0.230      0.204      0.219       A  No
A6   0.485       0.231      0.101      0.183       A  No
\end{verbatim}
\begin{figure}
\centering
\includegraphics[width = 0.9\textwidth]{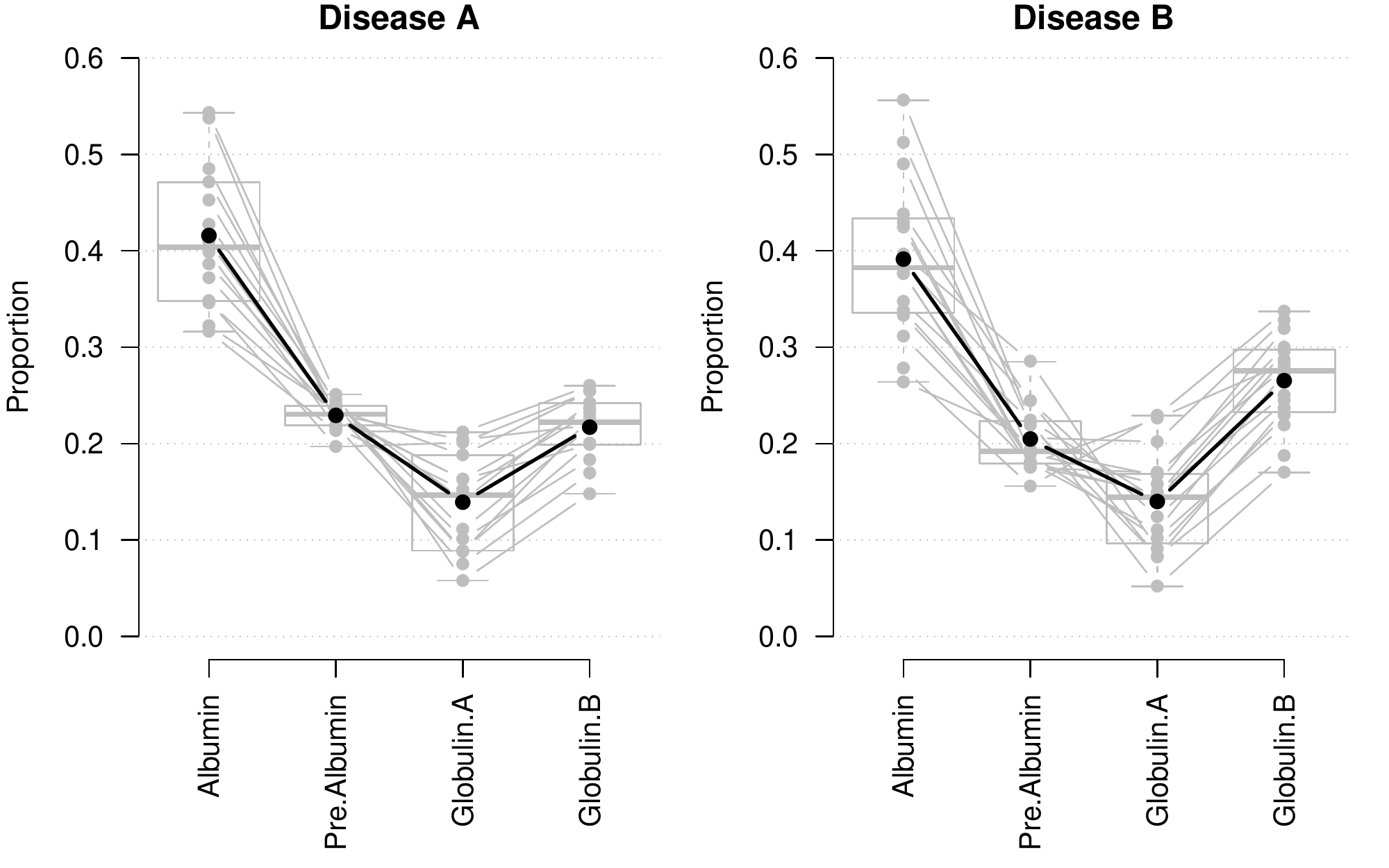}
\caption{Blood sample data: Grey points (lines indicate connections of proportions as coming from a shared observation unit) show the original data, black points the respective sample means.}
\label{fig:blood:samples:descriptive}
\end{figure}
Figure~\ref{fig:blood:samples:descriptive} on page~\pageref{fig:blood:samples:descriptive} visualizes the data. 
The task for a Dirichlet regression model is here to work out the differences in the composition of blood compartments by differences in the levels of a binary disease variable (\texttt{Disease}, unclassified observations are dropped). 

As $\mathbf{Y}$ distributes in $C\geq 2$ dimensions, ie.
$$
Y=\left\{\left(Y_1,\ldots, Y_C\right)^\top;\,Y_1+\ldots+Y_C=1\right\},
$$
one observation $\mathbf{y}_i=\left(y_{1,i},\ldots,y_{C,i}\right)^\top$ for observation unit $i$ is a vector of values $y_{c,i}$, $c=1, \ldots, C$, with $\sum_{i=1}^{C}y_{c,i}=1$. 
For each $Y_{c}$, one usually assumes a strictly positive component below $1$, ie. $Y_{c}\in\left(0, 1\right)\forall c\in\left\{1, \ldots, C\right\}$. 

\paragraph{Zero observations} In application, zero observations -- ie. $y_{c,i}=0$ -- are usually often present, either because the corresponding proportion $y_{c,i}$ is completely absent, or because $y_{c,i}$ is only observable with certain precision and consequently at risk to fall below a detection limit. 
Several strategies have been introduced of how to deal with, and how to distinguish among zero observations (see \cite{martinfernandezetal2003} for an overview). 
For the remainder of this article, we assume that any zero observation comes from the class of \textsl{rounded zeroes} \cite{martinfernandezetal2003}: 
\begin{quote}
"One typical example of rounded zero is the zero in a component of a particular mineral which indicates that no quantifiable proportion of the mineral has been recorded by the measurement process. 
Because this kind of zeros is usually understood as 'a trace too small to measure', it seems reasonable to replace them by a suitable small value [...]."
%\\
%\hfill 
\cite{martinfernandezetal2003}
\end{quote}
In case of zero observations, the transformation proposed by \cite{smithsonverkuilen2006} -- applied by \cite{maier2014} and implemented in \textsf{R} package \textsf{DirichReg} \cite{maier2015} in function \texttt{DR\_data} as argument \texttt{trafo} -- is applied:
\begin{align}
y_{c,i}=
\begin{cases}
y_{c,i},&\text{ if }y_{c,i}>0,\\
\dfrac{y_{c,i}\cdot\left(n-1\right)+\frac{1}{C}}{n},&\text{ else,}
\end{cases}
\end{align}
where $n$ is the number of observation units.
\paragraph{Dirichlet distribution} The Dirichlet distribution~\cite{bolshev2018} is defined as:
\begin{align}
f\left(\mathbf{y}\vert\boldsymbol{\alpha}\right)
=
\dfrac{1}{B\left(\boldsymbol{\alpha}\right)}\prod\limits_{c=1}^{C}y_{c}^{\alpha_{c}-1}
\end{align}
with shape parameter vector $\boldsymbol{\alpha}=\left(\alpha_1,\ldots,\alpha_C\right)^\top$, where the \textsl{multinomial beta function}, $B\left(\boldsymbol{\alpha}\right)=\left(\prod_{c=1}^{C}\Gamma\left(\alpha_{c}\right)\right)/\left(\Gamma\left(\prod_{c=1}^{C}\alpha_{c}\right)\right)$ serves as a normalizing constant, and $\Gamma\left(z\right)=\int_0^{\infty}t^{x-1}\exp\left(-t\right)\partial t$ denotes the \textsl{Gamma function}.

In the following, $\mathbf{Y}\sim D\left(\boldsymbol{\alpha}\right)$ will denote that a random variable $\mathbf{Y}=\left(Y_1,\ldots,Y_C\right)^\top$ is Dirichlet distributed. 
From this parametrisation, it is evident that the elements of $\boldsymbol{\alpha}$ must control location as well as scale of the distribution, where the sum $\alpha_0:=\sum_{c=1}^{C}\alpha_c$ of all $C$ components of parameter vector $\boldsymbol{\alpha}$ is interpreted as precision, ie. "the higher this value, the more density is near the expected value" \cite{maier2014}, while component-wise expected values are given by the ratio of parameters with precision, ie.
\begin{align}
E\left(Y_c\right)=\dfrac{\alpha_c}{\alpha_0}
\end{align}

Two direct consequences of this definition: 
\begin{itemize}
\item Expected values are proportions summing up to 1, and by this
\item the vector of $C$ expected values only has $C-1$ degrees of freedom: if we hold $C-1$ expected values fixed on certain values, the last expected value is given by the difference to 1. 
\end{itemize}
\paragraph{Regression modeling for Dirichlet distribution}
\cite{maier2014} introduces two parameter formulations for Dirichlet regression models, where here the so-called \textsl{alternative formulation} will be used:
\begin{align}
\mathbf{Y}_{i}\vert\mathbf{x}_i\sim D\left(\boldsymbol{\mu}_i,\theta\right),
\end{align}
with expectation vector $\boldsymbol{\mu}_i=\left(\mu_{1,i},\ldots,\mu_{C,i}\right)^\top$, expectation vector element $\mu_{c,i}=E\left(Y_{c,i}\vert\mathbf{x}_i\right)$, component index $c=1,\ldots,C$, and precision $\theta_i=\alpha_{0,i}$. % (with $\mu_c=\frac{\alpha_c}{\theta}$ we can translate back to the original formulation). 

Using component-wise coefficient vectors~$\boldsymbol{\beta}_c$ for components $c=1,\ldots,C$, we can set-up linear predictors $\eta_{c,i}$ defining $\mu_{c,i}$ as:
$$
\eta_{c,i}=\mathbf{x}_i^\top\boldsymbol{\beta}_c,
$$
%defining component-wise epected values 
where for one reference component $\tilde{c}$, all elements of $\boldsymbol{\beta}$ are equal to $0$, i.e. $\boldsymbol{\beta}_{\tilde{c}}=\mathbf{0}$, to guarantee identifiability as a consequence of the degrees of freedom reduced to $C-1$ (Usually, $\tilde{c}$ is selected as $1$ or $C$, here as $C$). 

Plugging these linear predictors into a ratio of model components -- after applying the exponential function guaranteeing positivity --, the component-wise expected values conditional on $\mathbf{x}_{i}$ are defined as:
$$
\mu_{c,i}=\dfrac{\exp\left(\eta_{c,i}\right)}{\sum\limits_{d=1}^{C}\exp\left(\eta_{i,d}\right)}.
$$
Applying again the exponential function, and introducing coefficient vector $\boldsymbol{\gamma}$, we can express the global precision conditional on $\mathbf{x}_{i}$ as:
$$
\theta_{i}=\exp\left(\mathbf{z}_i^\top\boldsymbol{\gamma}\right).
$$
We get to densities:
\begin{align}
%\scriptstyle
f\left(\mathbf{y}_{i}\vert\left(\mu_{i,1},\mu_{i,2},\ldots,\mu_{c,i}\right)^\top,\theta_{i}\right)=\nonumber\\=f\left(\mathbf{y}_{i}\vert\boldsymbol{\mu}_{i},\theta_{i}\right),
\end{align}
and likelihood:
\begin{align}
L\left(\left\{\mathbf{y}_i;\,i=1,\ldots,n\right\}\right)=\prod\limits_{i=1}^{n}
f\left(\mathbf{y}_{i}\vert\boldsymbol{\mu}_{i},\theta_{i}\right).
\end{align}
%% Why this article?
\paragraph{Why this manuscript?}
As seen above, one component is redundant since it is a linear combination of the others, but in practice we desire to know the relationship between the explanatory variables and the outcomes in all component. 
Often the components are equally important in the view of the researcher and it is not sensible to relegate one to reference status. 
The use of a Bayesian simulation framework allows us -- by the use of simple post-estimation calculations -- to quantify uncertainty in both the coefficient scales, as well as on the proportion scale. 
We are able to have a look on all quantities as an ensemble, get to statistical inference statements for all model components, and by this, are able to move between the 'x in c' and 'c across x' perspectives. %, as it easily allows 
%us to directly quantify uncertainty in both on the coefficients, and on the proportion scale. 
This is a great advantage in comparison to the ML framework \cite{maier2015}, where predictions are also easily achieved and readily implemented, but the expression of uncertainties for all quantities and/or on basis of proportions becomes quickly tedious, as handling variances and covariances is rather difficult for nonlinear combinations. 

\cite{vandermerwe2018} recently proposed a Bayesian solution to this practical limitation based on a penalized likelihood formulation respecting restrictions on the degrees of freedom as described above.
Following the simulation from the joint posterior, \cite{vandermerwe2018} applies post-simulation calculations "to ensure that fitted expected values sum to one for each observation" \cite{vandermerwe2018}:
$$
\mu_{c,i,S}^{\text{adj}}=\dfrac{\mu_{c,i,S}^{\text{sim}}}{\sum\limits_{d=1}^{C}\mu_{d,i,S}^{\text{sim}}}.
$$
This is a somewhat counter-intuitive step, as the sampling of coefficients is performed such that it should respect the degrees of freedom restriction by suitably penalizing the likelihood. 
Further, it introduces a certain removal of sampling variation that was initially accepted during the Bayesian sampling algorithm as being plausible with respect observed data and statistical model. 
By this one looses a direct linkage between the Bayesian model expressed by the joint posterior, and the adjusted samples from the joint posterior $\mu_{c,i,S}^{\text{adj}}$. 

In order to overcome this issue, but still be able to remain in the Bayesian framework that \cite{vandermerwe2018} introduced with good reason:
\begin{quote}
"Specifically, we use the Bayesian simulation framework, which holds many advantages. It allows us to directly quantify uncertainty in both the coefficients and the means. Also, when moving to a predictive framework, construction of predictive densities is relatively straightforward." \cite{vandermerwe2018}
\end{quote}
I will introduce a \textsf{Stan} implementation using one response component as reference status as used in the ML estimation framework by \cite{maier2014}, and demonstrate how post-estimation calculations overcome the limitations induced by reference status definition in the ML estimation framework. 
This post-estimation technique is simple: calculate the expected value $\mu_{C,i}$ -- conditional on the covariate values of an observation unit $i$ -- of reference component $C$ as the difference of the sum of the other expected values $\mu_1,\ldots,\mu_{C-1}$ to the boundary condition that all must sum up to 1:
$$
\mu_{C,i,s}=\dfrac{1}{\sum\limits_{d=1}^{C-1}\mu_{d,i,s}},
$$
-- this so far not different to what the we need to do to get to an estimate for $\mu_C$ in the ML framework -- repeatedly for each sample $s=1,\ldots,S$ from the posterior. 
By this we are equipped with not only one estimate, but a full distribution of samples for $\mu_C$ -- in the same fashion as for the other component -- and consequently, uncertainty measures for all response component can be expressed -- with, of course, however still reduced degrees of freedom of $C-1$. 

\cite{vandermerwe2018} based his software implementation on Gibbs sampling \cite{gelfandsmith1990} via the \textsf{OpenBUGS} program \cite{lunnetal2009}, but as I personally feel much more comfortable and experienced using the alternative Bayesian inference framework \textsf{Stan}, I will introduce the new approach using this alternative approach. 
However, any researcher should technically be able to perform her/his calculations in a similar way in the \textsf{OpenBUGS} framework. 

%% Overview remainder:
The remainder of this article is organized as follows: Section 2 introduces the Bayesian model as well as the \textsf{Stan} estimation framework for the Dirichlet regression model, and Section 3 shows an application example. 
% \newpage
\section{Methods and software implementation}
We move to a Bayesian framework and introduce normal priors with mean 0 and variance $5^2$ on all model parameters:
\begin{align}
\mathbf{y}_{i}\vert\mathbf{x}_i,\mathbf{z}_i&\sim D\left(\boldsymbol{\mu}_{i},\theta_{i}\right),\\
\mu_{c,i}&=\exp\left(\mathbf{x}_i^\top\boldsymbol{\beta}_c\right),\nonumber\\
\beta_{j,c}&\sim N\left(0,5^2\right),\quad j=1,\ldots,p_{\boldsymbol{\beta}},\quad \forall c\neq\tilde{c},\nonumber\\
\beta_{j,\tilde{c}}&=0,\quad j=1,\ldots,p_{\boldsymbol{\beta}},\nonumber\\
\gamma_{j,c}&\sim N\left(0,5^2\right),\quad j=1,\ldots,p_{\boldsymbol{\gamma}}.\nonumber
\end{align}
One can consider these normal priors as weakly-informative, as they certainly don't act uniformly on the coefficients, but as the informativeness of a prior does not depend on the absolute value per se, but rather on how flat it is in the region of high likelihood and vice versa, we can still consider that they let the data speak for themselves in almost any application with unit-scale covariates, as one can, for those unit-scale covariates, assume a region of high likelihood for coefficients in an interval of only a few units around 0.

\subsection{Software implementation}
The Bayesian estimation of additive regression models with Dirichlet distributed responses is implemented in \textsf{Stan}~\cite{carpenteretal2017} and practically applied using \textsf{RStan}~\cite{rstan2016}. 
\textsf{RStan} functions as an \textsf{R}~\cite{rcoreteam2017} interface to \textsf{Stan} using common \textsf{R} syntax. 
\textsf{Stan} is statistical software written in \textsf{C++} which operationalizes Bayesian inference by drawing samples from a regression model's joint posterior in the form of Markov chains. 
\textsf{Stan} is my personal generic Bayesian software framework -- it's a probabilistic programming language -- of choice as it is the current gold-standard~\cite{monnahanetal2017}, and I, the author, personally feel more comfortable and experienced with it in comparison to \textsf{JAGS}~\cite{plummer2003}, \textsf{Bugs}~\cite{gilksetal1994} and the like. 
\vskip 2mm
In contrast to the alternatives for Bayesian analyzes, \textsf{Stan} uses Hamiltonian Monte Carlo~\cite{duaneetal1987} and the No-U-Turn Sampler (NUTS)~\cite{hoffmangelman2014} which require fewer samples for chain convergence though at the cost of increased computation time per sample. 
% \vskip 2mm
In each of the further applications, a single model fit incorporates four chains with 2000 iterations each, where first 1000 iterations of each chain are discarded as warmup leaving in total 4000 post-warmup samples. 
These are defaults as for example also chosen by \textsf{brms}~\cite{buerkner2017}.
\vskip 2mm 
A \textsf{Stan} implementation for the Dirichlet regression model including covariate effects for the variables transported by \texttt{X} -- \textsf{R} base function \texttt{model.matrix(...)} is a convenient tool for preparation:
%\vskip 2mm 
%After running each model, the chains are assessed for convergence, where methods for checking chain convergence are:
%\begin{itemize}
%\item Log-posterior and posterior coefficient density plots are used to assess Markov Chain convergence. If the density estimates show clearly separated multiple modes during visual inspection, then the Markov Chains have not converged on a common coefficient domain. Essentially, if there were more than one mode, it would indicate the chains are still evaluating multiple points as plausible posterior regions. Across all models, the log posterior and posterior coefficients show no clear sign against unimodality indicating convergence and sufficient posterior sampling.
%\item Coefficient traceplots are another visual method of confirming convergence. The four chains for each model are plotted by variable.If the chains show trends of increasing or decreasing, or are not stationary, then they are concluded to have not converged. Across all models and variables, the chains appear to have converged indicating the selection of 1000 warmup samples and 1000 post-warmup samples is sufficient.  
%\item Coefficient $\hat{R}$, another measure of convergence, are also examined, where values greater than 1.1 indicate the chains have not sufficiently converged (Bürkner, 2017).
%\end{itemize}
\begin{verbatim}
stan_code <- '
data { 
  int<lower=1> N;  // total number of observations 
  int<lower=2> ncolY; // number of categories
  int<lower=2> ncolX; // number of predictor levels
  matrix[N,ncolX] X; // predictor design matrix
  matrix[N,ncolY] Y; // response variable
  real sd_prior; // Prior standard deviation
}
parameters {
  matrix[ncolY-1,ncolX] beta_raw; // coefficients (raw)
  real theta;
}
transformed parameters{
  real exptheta = exp(theta);
  matrix[ncolY,ncolX] beta; // coefficients
  for (l in 1:ncolX) {
    beta[ncolY,l] = 0.0;
    }
  for (k in 1:(ncolY-1)) {
    for (l in 1:ncolX) {
      beta[k,l] = beta_raw[k,l];
      }
    }
  }
model {
  // prior:
  theta ~ normal(0,sd_prior);
  for (k in 1:(ncolY-1)) {
    for (l in 1:ncolX) {
      beta_raw[k,l] ~ normal(0,sd_prior);
      }
    }
  // likelihood 
  for (n in 1:N) {
    vector[ncolY] logits;
    for (m in 1:ncolY){
      logits[m] = X[n,] * transpose(beta[m,]);
      }
    transpose(Y[n,]) ~ dirichlet(softmax(logits) * exptheta); 
    }
  }
'
\end{verbatim}
Supplement S1 gives an implementation of alternative parametrization \cite{maier2014} Dirichlet regression including varying precision by varying covariates. 
%\newpage
\section{Results for the Blood Samples Application}
\begin{figure}
\centering
\includegraphics[width = 0.9\textwidth]{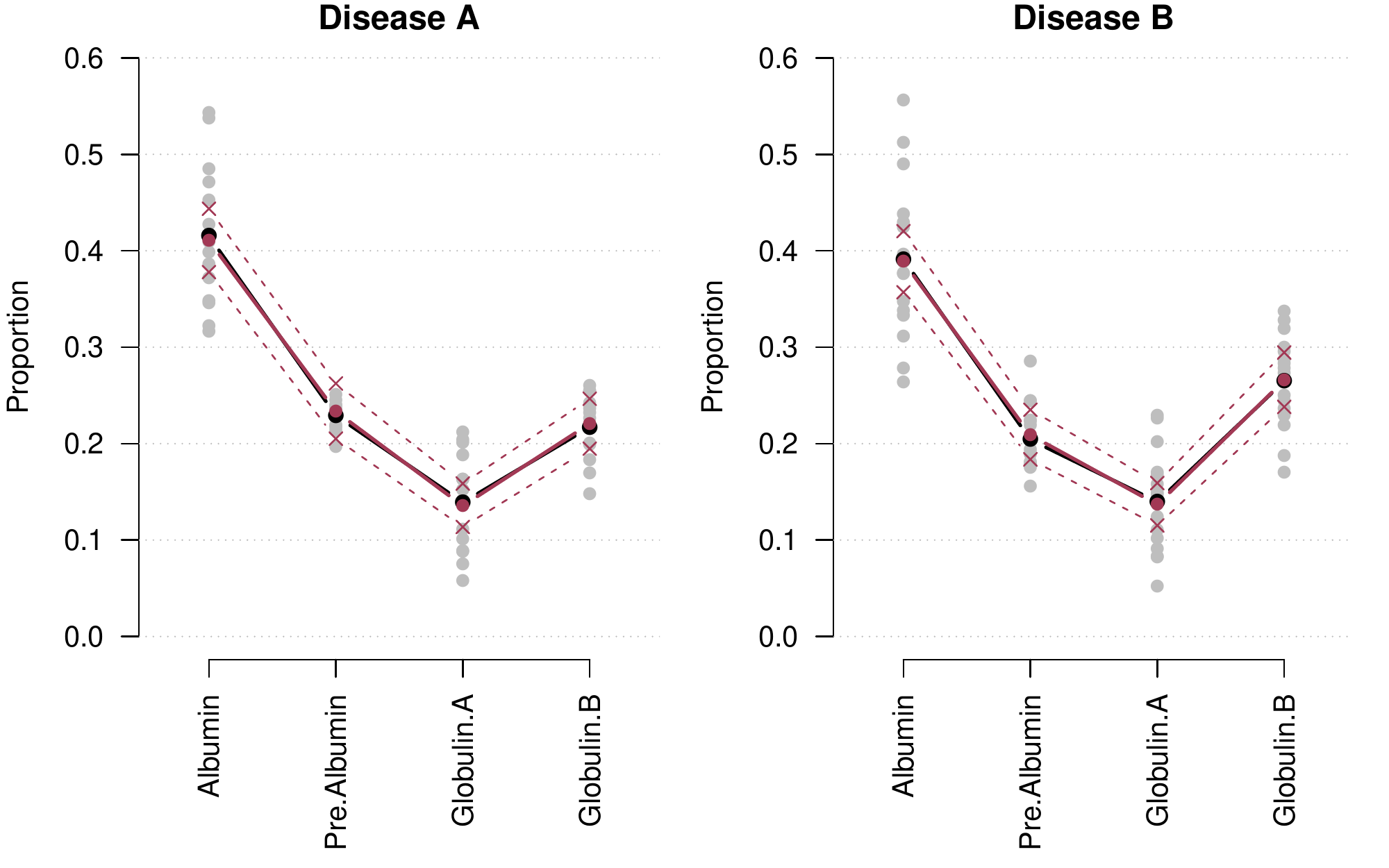}
\caption{Results for the blood sample data: Grey points show the original data, black points show the respective sample mean, red points give the estimated expected values conditional on disease based on the Bayesian approach, and black crosses show 95\% credible (point-wise) intervals based on the Bayesian estimation approach.}
\label{fig:blood:samples:results}
\end{figure}

First, let's see how these data is analyzed using \textsf{DirichletReg}:
%Blood samples (See section 4.2 in Maier 2014): We need this application to proof the concept for categorical covariates as this will be the scenario in the application at NW-FVA; I have already done that on the real application example and there are only numerical differences in uncertainty (confidence/credible) intervals and point (ML/Posterior mean) estimates, so that works! (But as the data will published there, the blood sample example will have to do the job at the moment)!
\vskip 2mm
\begin{verbatim}
Bld <- na.omit(Bld)
Bld$Smp <- DR_data(Bld[, 1:4])
\end{verbatim}
We get a warning that some values needed to be corrected such that all proportions for one observation unit sum up to 1:
\begin{verbatim}
Warning in DR_data(Bld[, 1:4]) :
  not all rows sum up to 1 => normalization forced
\end{verbatim}
The ML estimate is calculate via:	
\begin{verbatim}
m <- DirichReg(Smp ~ Dis. | 1, Bld, model = "alternative", base = 4)
\end{verbatim}
We get the following coefficient estimates:
\begin{verbatim}
(b <- unlist(coef(m)))
         beta.Alb..(Interc.)              beta.Alb..Dis.B 
                  0.63010700                  -0.25191609 
     beta.Pre.Alb..(Interc.)          beta.Pre.Alb..Dis.B 
                  0.06274025                  -0.30952737 
       beta.Glob.A.(Interc.)            beta.Glob.A.Dis.B 
                 -0.48628655                  -0.18189666 
       gamma.gamma.(Interc.) 
                  4.22272495 
\end{verbatim}
And can calculate expected values by:
\begin{verbatim}
(B <- matrix(nrow = 4, ncol = 2, c(b[1:6], rep(0, 2)), byrow = T))
            [,1]       [,2]
[1,]  0.63010700 -0.2519161
[2,]  0.06274025 -0.3095274
[3,] -0.48628655 -0.1818967
[4,]  0.00000000  0.0000000
(eB <- t(exp(apply(B, MAR = 1, FUN = cumsum))))
          [,1]      [,2]
[1,] 1.8778115 1.4596416
[2,] 1.0647502 0.7813070
[3,] 0.6149056 0.5126391
[4,] 1.0000000 1.0000000
(mu <- cbind(eB[, 1]/colSums(eB)[1], eB[, 2]/colSums(eB)[2]))
          [,1]      [,2]
[1,] 0.4120296 0.3888657
[2,] 0.2336276 0.2081494
[3,] 0.1349227 0.1365731
[4,] 0.2194201 0.2664118
\end{verbatim}
\vskip 2mm
Now using the Bayesian estimation using \textsf{Stan} begins with compiling the previously introduced \textsf{Stan} code:
\begin{verbatim}
library("rstan")
prg <- stan_model(model_code = stan_code)
\end{verbatim}
We then need to translate the covariates into a design matrix \texttt{X}:
\begin{verbatim}
X <- as.matrix(model.matrix(lm(Albumin ~ Disease, data = Bld)))
X <- matrix(nrow = nrow(X), ncol = ncol(X), data = as.numeric(X))
\end{verbatim}
Define response object \texttt{Y}:
\begin{verbatim}
Y <- Bld$Smp
\end{verbatim}
Provide everything as a list:
\begin{verbatim}
D <- list(N = nrow(Y), ncolY = ncol(Y), ncolX = ncol(X),
          X = X, Y = Y, sd_prior = 1)
\end{verbatim}
And finally estimate the parameters using function \texttt{sampling} from the \textsf{RStan} package:
\begin{verbatim}
fit1 <- sampling(prg, data = D, chains = 4, iter = 2000, cores = 4,
                 control = list(adapt_delta = 0.95, max_treedepth = 20), 
                 refresh = 100)
\end{verbatim}
Using \texttt{extract}, we can access the posterior samples:
\begin{verbatim}
B <- extract(fit1)$beta
\end{verbatim}
A small helping-function will assist us with calculating the expected values:
\begin{verbatim}
simplex <- function(x){
  exp(x)/sum(exp(x))
}
\end{verbatim}
We can plot everything using the following code:
\begin{verbatim}
plot(1:4, Bld[1, 1:4], ylim = c(0, 0.6), type = "n", xaxt = "n", las = 1, 
     xlab = "", ylab = "Proportion", main = "Disease A", xlim = c(0.6, 4.4))
abline(h = seq(0, 0.6, by = 0.1), col = "grey", lty = 3)
axis(1, at = 1:4, labels = names(Bld)[1:4], las = 2)
aux <- t(apply(B[, , 1], MAR = 1, FUN = simplex))
apply(subset(Bld, Disease == "A")[, 1:4], MAR = 1, FUN = points, pch = 16, 
      col = "grey")
lines(apply(subset(Bld, Disease == "A")[, 1:4], MAR = 2, FUN = mean), 
      type = "b", pch = 16, cex = 1.2, lwd = 2)
lines(apply(aux, MAR = 2, FUN = quantile, prob = 0.975), type = "b", pch = 4, 
      lty = 2, col = my_colors[2])
lines(apply(aux, MAR = 2, FUN = quantile, prob = 0.025), type = "b", pch = 4, 
      lty = 2, col = my_colors[2])
lines(apply(aux, MAR = 2, FUN = mean), lwd = 2, col = my_colors[2], type = "b", 
      pch = 16)
plot(1:4, Bld[1, 1:4], ylim = c(0, 0.6), type = "n", xaxt = "n", las = 1, 
     xlab = "", ylab = "Proportion", main = "Disease B", xlim = c(0.6, 4.4))
abline(h = seq(0, 0.6, by = 0.1), col = "grey", lty = 3)
axis(1, at = 1:4, labels = names(Bld)[1:4], las = 2)
aux <- t(apply(B[, , 1] + B[, , 2], MAR = 1, FUN = simplex))
apply(subset(Bld, Disease == "B")[, 1:4], MAR = 1, FUN = points, pch = 16, 
      col = "grey")
lines(apply(subset(Bld, Disease == "B")[, 1:4], MAR = 2, FUN = mean), 
      type = "b", pch = 16, cex = 1.2, lwd = 2)
lines(apply(aux, MAR = 2, FUN = quantile, prob = 0.975), type = "b", pch = 4, 
      lty = 2, col = my_colors[2])
lines(apply(aux, MAR = 2, FUN = quantile, prob = 0.025), type = "b", pch = 4, 
      lty = 2, col = my_colors[2])
lines(apply(aux, MAR = 2, FUN = mean), lwd = 2, col = my_colors[2], type = "b", 
      pch = 16)
\end{verbatim}
Figure~\ref{fig:blood:samples:results} visualizes the outcome of these \textsf{R} plotting commands. 
%% On this scale, measures of uncertainty are not easily accessible using the ML approach. 
Note that by the reduced degrees of freedom, results should be seen only in the ensemble as the outcome of one proportion influences all the other proportions' outcomes, and vice versa.
 
Table~\ref{result:table} shows the results on the parameter level and compares ML to Bayes for several different prior choices. As can be seen, different prior choices only have an practically irrelevant influence on the parameters' posterior distribution. 

\vskip 2mm
\begin{table}
\centering
\begin{minipage}[l]{8cm}
Maximum Likelihood:$\vphantom{N(0,1^2)}$
\\
\begin{tabular}{rrrr}
\hline
& 2.5\% & Est. & 97.5\% \\ 
\hline
Alb.:(Interc.) & 0.46 & 0.63 & 0.80 \\ 
Alb.:Dis.B & -0.47 & -0.25 & -0.03 \\ 
Pre.Alb.:(Interc.) & -0.12 & 0.06 & 0.25 \\ 
Pre.Alb.:Dis.B & -0.56 & -0.31 & -0.06 \\ 
Glob.A:(Interc.) & -0.70 & -0.49 & -0.27 \\ 
Glob.A:Dis.B & -0.47 & -0.18 & 0.11 \\ 
Glob.B:(Interc.) & 0.00 & 0.00 & 0.00 \\ 
Glob.B:Dis.B & 0.00 & 0.00 & 0.00 \\ 
\hline
\end{tabular}
%\end{table}
\end{minipage}
\begin{minipage}[l]{8cm}
Bayes with $N(0,1^2)$ prior:
\\
\begin{tabular}{rrrr}
\hline
& 2.5\% & Est. & 97.5\% \\ 
\hline
Alb.:(Interc.) & 0.45 & 0.62 & 0.80 \\ 
Alb.:Dis.B & -0.47 & -0.24 & -0.00 \\ 
Pre.Alb.:(Interc.) & -0.14 & 0.06 & 0.26 \\ 
Pre.Alb.:Dis.B & -0.57 & -0.30 & -0.03 \\ 
Glob.A:(Interc.) & -0.71 & -0.49 & -0.26 \\ 
Glob.A:Dis.B & -0.48 & -0.18 & 0.13 \\ 
Glob.B:(Interc.) & 0.00 & 0.00 & 0.00 \\ 
Glob.B:Dis.B & 0.00 & 0.00 & 0.00 \\ 
\hline
\end{tabular}
%\end{table}
\end{minipage}
\vskip 2mm
\begin{minipage}[l]{8cm}
Bayes with $N(0,5^2)$ prior:
\\
%\begin{table}[ht]
%\centering
\begin{tabular}{rrrr}
\hline
& 2.5\% & Est. & 97.5\% \\ 
\hline
Alb.:(Interc.) & 0.46 & 0.63 & 0.80 \\ 
Alb.:Dis.B & -0.48 & -0.25 & -0.02 \\ 
Pre.Alb.:(Interc.) & -0.14 & 0.06 & 0.25 \\ 
Pre.Alb.:Dis.B & -0.57 & -0.30 & -0.05 \\ 
Glob.A:(Interc.) & -0.72 & -0.49 & -0.27 \\ 
Glob.A:Dis.B & -0.48 & -0.18 & 0.12 \\ 
Glob.B:(Interc.) & 0.00 & 0.00 & 0.00 \\ 
Glob.B:Dis.B & 0.00 & 0.00 & 0.00 \\ 
\hline
\end{tabular}
%\end{table}
\end{minipage}
\begin{minipage}[l]{8cm}
Bayes with $N(0,50^2)$ prior:
\\
%\begin{table}[ht]
%\centering
\begin{tabular}{rrrr}
\hline
& 2.5\% & Est. & 97.5\% \\ 
\hline
Alb.:(Interc.) & 0.46 & 0.63 & 0.80 \\ 
Alb.:Dis.B & -0.48 & -0.25 & -0.02 \\ 
Pre.Alb.:(Interc.) & -0.13 & 0.06 & 0.26 \\ 
Pre.Alb.:Dis.B & -0.58 & -0.31 & -0.05 \\ 
Glob.A:(Interc.) & -0.71 & -0.49 & -0.27 \\ 
Glob.A:Dis.B & -0.48 & -0.18 & 0.11 \\ 
Glob.B:(Interc.) & 0.00 & 0.00 & 0.00 \\ 
Glob.B:Dis.B & 0.00 & 0.00 & 0.00 \\ 
\hline
\end{tabular}
\end{minipage}
\caption{Results for Maximum Likelihood and Bayesian estimation in the blood sample data on the parameter level: Comparison between ML and Bayes with several different prior choices.}
\label{result:table}
\end{table}

\clearpage
\section{Discussion}
We have introduced an implementation of a Bayesian estimation framework for Dirichlet regression that combines the advantage of identifiability -- by selection of a reference category -- with post-estimation flexibility -- by use of a Bayesian simulation algorithm. 
Directly to be used \textsf{Stan} code was introduced, and an application demonstrates the modeling capabilities of this solution. 
%\clearpage
\bibliography{literatur}{}
\bibliographystyle{plain}
\clearpage
\beginsupplement
\section{\textsf{Stan} implementation of the alternative parametrization including varying precision}
The following code uses design matrix object \texttt{X} both for expectation and precision:
\begin{verbatim}
stan_code_var_theta <- '
data {
int<lower=1> N;  // total number of observations
int<lower=2> ncolY;  // number of categories
int<lower=2> ncolX; // number of predictor levels
matrix[N,ncolX] X; // predictor design matrix
matrix[N,ncolY] Y;  // response variable
real sd_prior_beta; // Prior standard deviation
real sd_prior_theta; // Prior standard deviation
}
parameters {
matrix[ncolY-1,ncolX] beta_raw; // coefficients (raw)
matrix[1,ncolX] theta; // coefficients
}
transformed parameters{
matrix[ncolY,ncolX] beta; // coefficients
for (l in 1:ncolX) {
beta[ncolY,l] = 0.0;
}
for (k in 1:(ncolY-1)) {
for (l in 1:ncolX) {
beta[k,l] = beta_raw[k,l];
}
}
}
model {
// priors
for (l in 1:ncolX) {
for (k in 1:(ncolY-1)) {
beta_raw[k,l] ~ normal(0,sd_prior_beta);
}
theta[1,l] ~ normal(0,sd_prior_theta);
}
// likelihood
for (n in 1:N) {
vector[ncolY] logits;
real exptheta;
for (m in 1:ncolY){
logits[m] = X[n,] * transpose(beta[m,]);
}
exptheta = exp(X[n,] * transpose(theta[1,]));
logits = softmax(logits);
transpose(Y[n,]) ~ dirichlet(logits * exptheta);
}
}
'
\end{verbatim}
\end{document}